\documentclass[12pt,preprint]{aastex}

\newcommand{\ang}{\AA\hspace{.01cc} } 

\shorttitle{DENIS J081730.0-615520: An overlooked mid-T dwarf in the solar neighborhood}
\shortauthors{Artigau et al.}

\begin{document}

\title{DENIS J081730.0-615520: An overlooked mid-T dwarf in the solar neighborhood}

\author{\'Etienne Artigau\altaffilmark{1}, Jacqueline Radigan\altaffilmark{2}, Stuart Folkes\altaffilmark{3}, Ray Jayawardhana\altaffilmark{2}, Radostin Kurtev\altaffilmark{3}, David Lafreni\`ere\altaffilmark{1}, Ren\'e Doyon\altaffilmark{1}, Jura Borissova\altaffilmark{3}}

\altaffiltext{1}{D\'epartement de Physique and Observatoire du Mont M\'egantic, Universit\'e de Montr\'eal, C.P. 6128, Succ. Centre-Ville, Montr\'eal, QC, H3C 3J7, Canada}
\altaffiltext{2}{Department of Astronomy and Astrophysics, University of Toronto, 50
St. George Street, Toronto, ON, M5S 3H4, Canada}
\altaffiltext{3}{Departamento de F\'isica y Astronom\'ia, Facultad de Ciencias, Universidad de
Valpara\'iso, Ave. Gran Breta\~na 1111, Playa Ancha, Casilla 53, Valpara\'iso, Chile}

\email{artigau@astro.umontreal.ca}

\begin{abstract}
Recent wide-field near-infrared surveys have uncovered a large number of cool brown dwarfs, extending the temperature sequence down to less than 500\,K and constraining the faint end of the luminosity function. One interesting implication of the derived luminosity function is that the brown dwarf census in the immediate ($<10$\,pc) solar neighborhood is still largely incomplete, and some bright ($J<16$) brown dwarfs remain to be identified in existing surveys. These objects are especially interesting as they are the ones that can be studied in most detail, especially with techniques that require large fluxes (e.g. time-variability, polarimetry, high-resolution spectroscopy) that cannot realistically be applied to objects uncovered by deep surveys. By cross-matching the DENIS and the 2MASS point-source catalogs, we have identified an overlooked brown dwarf ---DENIS J081730.0-615520--- that is the brightest field mid-T dwarf in the sky ($J = 13.6$). We present astrometry and spectroscopy follow-up observations of this brown dwarf. Our data indicate a spectral type T6 and a distance --from parallax measurement-- of $4.9\pm0.3$\,pc, placing this mid -T dwarf among the 3 closest isolated brown dwarfs to the Sun.

\end{abstract}
\noindent{\em Suggested running page header:} 

\section{Introduction}
Recent, deep wide-field near-infrared surveys (Canada France Brown Dwarf Survey, \citet{Delorme2008b}; UKIRT Infrared Deep Sky Survey, \citet{Gonzalez2010}; CFBDS, UKIDSS) have uncovered large samples of faint brown dwarfs (BDs) and extended the cool end of the temperature sequence of known objects from 800\,K to 500\,K. These deep surveys provide the best estimate of the BD luminosity function (LF) to date and the results indicate that the sample of known, bright ($J<16$), brown dwarfs in the solar neighborhood is incomplete. For example, the LF derived from the CFBDS indicates that there are twice as many T dwarfs per unit volume as there are currently known in the solar neighborhood (${\rm d}<10$\,pc, \citealt{Reyle2009}).

There is ample evidence that brown dwarf searches based on Sloan Digital Sky Survey (SDSS), DEep Near-Infrared Survey of the Southern sky (DENIS) and the Two-Micron All Sky Survey (2MASS) are incomplete. For example, the recent discoveries of the brightest early-T dwarf, SIMP\,J0136+09 ($J=13.5$ \citealt{Artigau2006}), a bright L/T transition object, 2MASS\,J1126‚àí50 ($J=14.0$, \citealt{Folkes2007}), and a very bright blue L dwarf with a late-T companion, SDSS\,J1416+13AB  ($J=13.1$, \citealt{Bowler2010, Schmidt2010, Burningham2010}), all within $\sim8$ pc from the Sun, show that objects up to 3 magnitudes brighter than the completeness limits of all-sky surveys still remain to be discovered.

Identifying the nearest brown dwarfs is especially important as they can be observed with techniques that cannot be applied to the majority of objects uncovered by large, deep, surveys, such as time-variability, polarimetry or high-resolution spectroscopy. Bright objects further serve as references to understand similar but fainter objects.

\section{Discovery of DENIS J081730.0-615520}

 We recently undertook a cross-match of DENIS \citep{Epchtein1997} $I$-band dropouts with 2MASS \citet{Cutri2003, Skrutskie2006} point source catalog to search for overlooked high-proper motion brown dwarfs. The search was not limited in galactic latitude but in stellar density, dividing the DENIS catalog in $2^{\circ}\times2^{\circ}$ boxes and avoiding the $20\%$ of fields with the highest stellar density. The cutoff density corresponds to $\sim11$ DENIS sources per square arcminute. This effectively removed an elliptical region of $\pm15^{\circ}$ in galactic latitude and $\pm90^{\circ}$ in galactic longitude and areas around both Magellanic Clouds. We selected all DENIS $I$-band dropouts that were matched to a 2MASS source that was itself unmatched to a visible object ({\it ass\_opt} keyword in the 2MASS PSC) and at an angular distance consistent with a proper motion between $0.5\arcsec$ yr$^{-1}$ and $10\arcsec$ yr$^{-1}$. The best candidate identified so far, DENIS J081730.0-615520 (thereafter DENIS0817) has a proper motion of $\sim$1.2$\arcsec$ yr$^{-1}$ and near-infrared colors of $J-H=0.08$, $H-K=0.01$ and $I-J>5.9$ at the 1\,$\sigma$ limit, indicative of a nearby T5-T7 BD. The brightness of this object ($J=13.6$) made it a high priority for follow-up observations.

\begin{figure}[!htb]
\epsscale{1.02}
\plotone{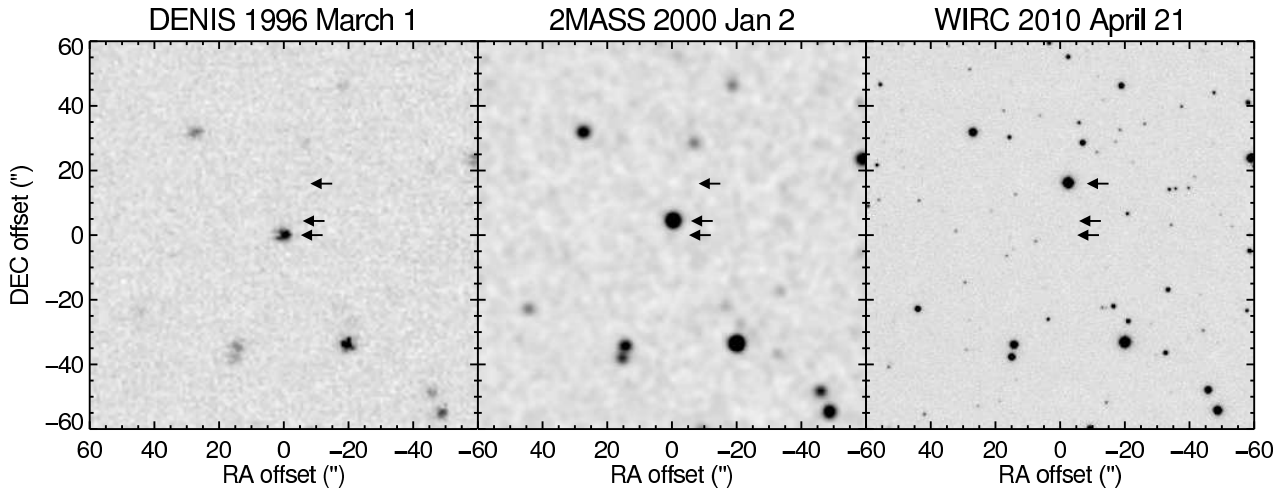}
\caption{\label{fig1}$J$-band DENIS, 2MASS and WIRC images of DENIS0817 showing its large proper motion relative to background objects. The follow-up confirmation WIRC image was used to further constrain the proper motion and the parallax.}
\end{figure}

\section{Las Campanas imaging and astrometry }\label{imaging}

\begin{figure*}[!tbh]
\epsscale{.99}
\plotone{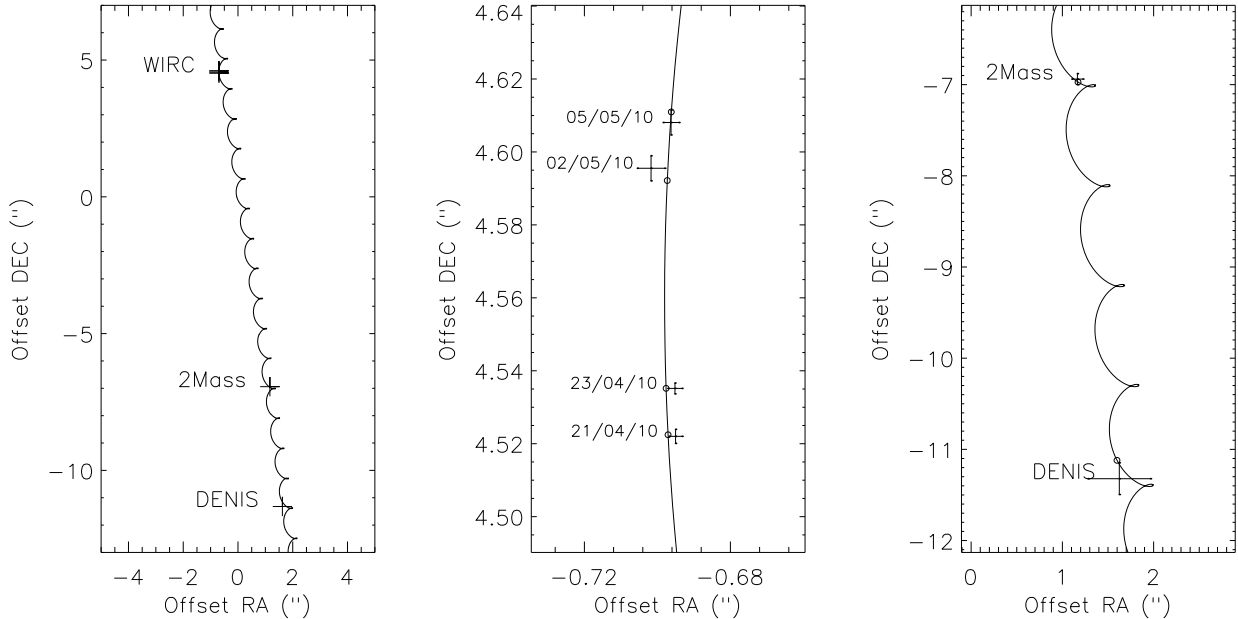}
\caption{\label{fig2}Astrometric measurements from the DENIS $J$-band, 2MASS and WIRC observations. Left panel shows all measurements at hand and the fitted astrometric solution; individual WIRC measurements cannot be resolved individually in this plot. The DENIS and 2MASS versus the mean WIRC position put strong constraints on the proper motion, while the difference between the measured motion within the WIRC dataset and the mean proper motion constraints parallax. Central panel shows the WIRC dataset only; the time of year of these observations was such that proper motion and parallax each account for half of the sky motion of DENIS0817. Right panel shows the DENIS and 2MASS positions. For the central and left panels, the open circles show the position given by the astrometric and parallax fit for each epoch of observation.}
\end{figure*}

Following the identification of DENIS0817, $J$-band imaging was obtained at the DuPont 2.5-m telescope at Las Campanas using the Wide-field InfraRed Camera (WIRC, \citealt{Persson2002}) on 2010 April 21. A sequence of 18 dithered images was taken with a 30-s exposure time. The images were sky-subtracted and flat-fielded. As these follow-up images detected the source and confirmed the proper motion (see Figure \ref{fig2}) deduced from the DENIS and 2MASS images, 3 additional sequences were taken on April 23, May 3 and May 5 with at least 10 images each. An astrometric solution was found for every epoch using 3 stars near DENIS0817 (2MASS08172725-6155537, 2MASS08171910-6155227, 2MASS08173928-6154221). These stars have small proper motions ($<15$\,mas yr${^{-1}}$ with uncertainties of $\sim5$\,mas yr$^{-1}$; \citealt{Zacharias2005}) that have been taken into account in  calculating the astrometric solution. For each epoch, the mean of the positions measured in individual frames was taken and its uncertainty determined from the dispersion of values in each spatial direction. Figure \ref{fig2} shows the positions in DENIS (1996 March 1), 2MASS (2000 January 2) and WIRC images (2010 April 21 -- May 5). A parallax and proper motion fit was done on the astrometric measurements (see Table \ref{tbl-1} and Figure \ref{fig2}); the $\chi^{2}=0.98$ for the fit indicates that the error estimates are realistic.

The WIRC observations on 2010 April 23 were taken as part of an L and T dwarf variability program. The target was monitored in the $J$-band over a 4 hour baseline using 45\,s exposures and a random dither sequence within a 75 pixel radius.  Conditions were generally clear, with intermittent light cirrus, and $0.8\arcsec$ seeing.  Our lightcurve of DENIS0817 shows no evidence for variability above the $1$\% level. 

\section{Spectroscopy}\label{spectro}
Near-infrared spectroscopy of DENIS0817 was obtained on 2010 May 2 with the Ohio State Infrared Imager/Spectrometer (OSIRIS; \citealt{DePoy1993}), attached to the SOAR 4.1\,m telescope located on Cerro Pach{\'o}n, Chile. The spectrograph was used in the cross-dispersed mode using a slit of $1\arcsec$ width (0.34\,mm) with a plate scale of 0.331$\arcsec$ pixel$^{-1}$ (F2.8), giving a resolving power of R $\sim 1200$ for all {\em J}, {\em H}, and {\em K} spectral bands (orders $\mbox{n} = \mbox{3 -- 5}$). DENIS0817 was observed at the beginning of the night with the seeing for all observations between 1 -- 1.4$\arcsec$. A sequence of eight dithered 60-s exposures were taken, as well as a further four separate science-- sky pairs, also with exposures of 60\,s each, giving a total of on-target integration time of 720\,s. For the calibration frames a series of 20 flat fields (both for lamp on and off) using the internal instrument lamp, as well as argon/neon arc frames, were taken before the science frames. Immediately after the science frames the A0 star HIP43796 was observed at a similar airmass ($\mbox{difference} < 0.1$) to correct for telluric and instrumental transmission.

\begin{figure*}[!htbp]
\epsscale{.99}
\plotone{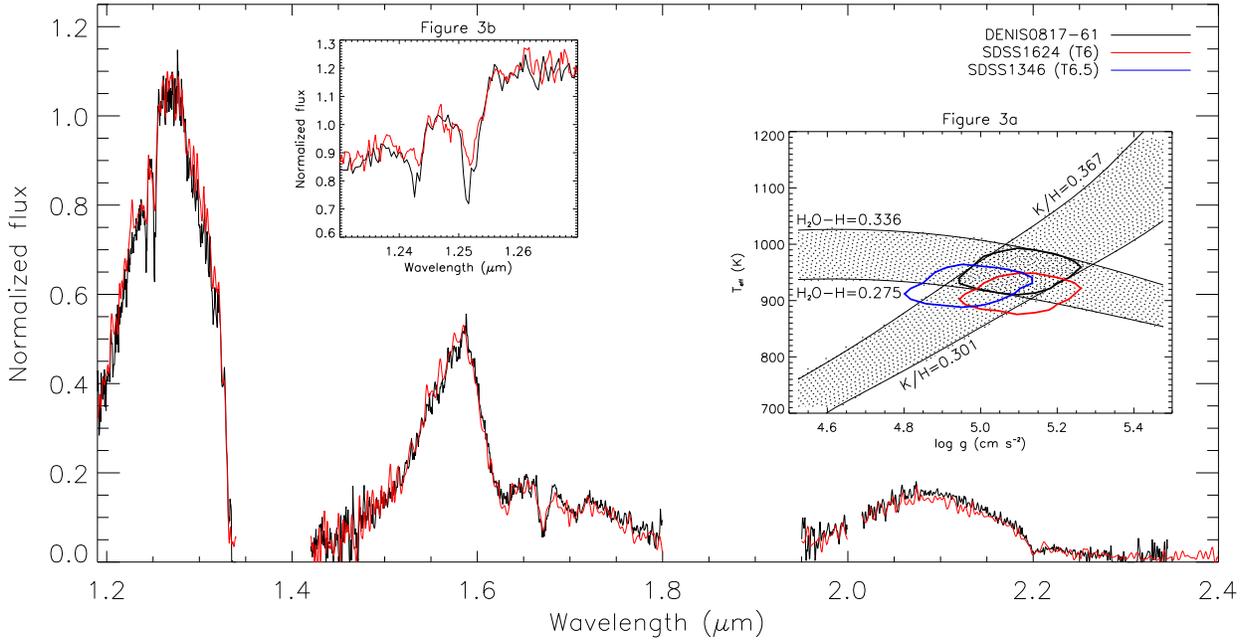}
\caption{\label{fig3} Spectra of DENIS0817 and SDSS1624+00. The T$_{\rm eff}$ versus $\log g$ inset (Figure 3a) shows the parameter space constrained by the H$_2$O-H and K/H indexes for DENIS0817 (dotted area and black line), SDSS1624+00 (red line) and SDSS1346-00 (blue line). The  K{\sc I} doublet inset (Figure 3b) highlights the difference in the depth of this doublet between these two objects. The SDSS1624+00 spectrum in the inset is from the \citet{McLean2003} sample while the low-resolution one spawning the whole near-infrared interval is taken from \citet{Strauss1999}. }
\end{figure*}

The dataset was reduced by pair-subtracting consecutive science images, dividing by the median-combined flat frame, extracting the four cross-dispersed orders and correcting for order curvature. The spectra of both the telluric and science target were extracted using Gaussian-weighted extraction boxes on both the positive and negative traces. All individual spectra were normalized and median combined in a single spectrum. Wavelength calibration was done using the argon arc spectrum. The atmosphere and instrument transmission function was determined by dividing the measured telluric star spectrum by a ${\rm T}=9900$\,K black body; the hydrogen series lines were then fitted and subtracted. The cross-dispersed orders were combined by normalizing their overlapping wavelength intervals. The $K$-band order overlaps with the $H$-band only in the deep water absorption around 1.9\,$\mu$m where continuum measurement is challenging; the $H/K$ normalization was therefore done using the 2MASS photometry. The synthesized $J-H$ color, as derived from the spectrum, matches the 2MASS color to $1\%$, well within the photometric uncertainties, and was not corrected. The useful wavelength interval spans 1.18\,$\mu$m to 2.30\,$\mu$m. Figure \ref{fig3} shows the spectrum of DENIS0817 and Table \ref{tbl-1} lists the spectral indices derived from this spectrum.

\section{Discussion}\label{discu}

The DENIS0817 spectrum closely matches that of SDSS1624+00, the standard for the T6 spectral type \citep{Strauss1999,Burgasser2006}. To constrain quantitatively the physical differences between these two objects, we reproduced the \citet{Burgasser2006e} analysis to constrain $\log g$ and T$_{\rm eff}$. As the H$_2$O-J index used in  \citet{Burgasser2006e} is not covered by our near-infrared spectrum, we used the H$_2$O-H versus K/H indexes instead of H$_2$O-J versus K/H. This analysis was performed for DENIS0817, SDSS1624+00 and SDSS1346-00 (T6.5, \citealt{Tsvetanov2000}). The $\log g$ and T$_{\rm eff}$ derived for the two latter objects differ slightly from those obtained by  \citet{Burgasser2006e} using H$_2$O-J and K/H ($\sim$80\,K cooler), but the differences in T$_{\rm eff}$ and $\log g$ between the objects are consistent with those derived by \citet{Burgasser2006e}. Note that \citet{DelBurgo2009} analyzed high-resolution, near-infrared spectra of a sample of mid and late-Ts that include SDSS1346 and SDSS1624 and found qualitatively the same difference between these objects, although they derived $\log g$ values $\sim 0.85$\,dex lower than those of \citet{Burgasser2006e}. From the region constrained in Figure \ref{fig3}a, DENIS0817 is found to have the same $\log g$ as SDSS1624 (within $\sim$$0.1$\,dex) and a warmer temperature by $40\pm40$\,K. As shown in Figure \ref{fig3}b, the 1.24$\mu$m K{\sc I} doublet is significantly deeper for DENIS0817 than it is for SDSS1624. For earlier spectral types, a deep K{\sc I} doublet is an indication of lower surface gravity, but the T6 spectral type corresponds to a rapid drop in the depth of this feature with temperature (see both panels of Figure 15 in \citealt{McLean2003}) and, considering the position of SDSS1624+00 and DENIS0817 in the $\log g$ versus T$_{\rm eff}$ diagram, we interpret the difference in the depth of the doublet as a signature of the slightly cooler temperature of SDSS1624+00 rather than a significant difference in surface gravity. Note that the measured values of equivalent widths for the KI[$1.243\mu$m] and KI[$1.254\mu$m] lines are slightly lower than the plateau in equivalent widths between T0 and T5 in \citet{McLean2003} ($\sim5$\ang for KI[$1.243\mu$m] and $\sim8$\ang for KI[$1.254\mu$m]) suggesting that DENIS0817 is just at the temperature where the doublet first begins to rapidly weaken with decreasing temperature. Thus, the KI doublet cannot be reliably used to constrain surface gravity or metallicity. The presence of the KI doublet does exclude the possibility that this object is metal poor, such as 2MASS 0937+2931 \citep{Burgasser2003b, Burgasser2006e}. This is consistent with its kinematics; the $27\pm2$\,km s$^{-1}$ tangential motion of DENIS0817 is typical of field BDs and indicative of a thin disk membership \citep{Faherty2009}.

The measured distance for DENIS0817 is $4.9\pm0.3$\,pc, making it one of the closest BDs to the Sun, and only second (or third, given the uncertainty) among isolated BDs, the closest previously known being UGPS J0722-05 (T9+, $2.9\pm0.4$ pc, \citealt{Lucas2010}) and DENIS-P J0255-4700, (L8, $5.0\pm0.1$ pc, \citealt{Martin1999, Costa2006}). There are also 3 T dwarfs in orbit around nearby stars within 5 pc: $\epsilon$ Indi Bab (T1/T6, $3.63\pm0.01$\,pc, \citealt{Scholz2003}) and SCR 1845-6357B (T6, $3.85\pm0.02$\,pc, \citealt{Biller2006}).

DENIS0817 (M$_J[\rm MKO]$ $= 14.85\pm0.14$) is slightly brighter than SDSS1624+00 (M$_J[\rm MKO] = 14.99$); this difference is consistent with their $\sim40$\,K temperature difference. Taking the derivative of the \citet{Knapp2004} M$_J$ versus spectral type and \citet{Golimowski2004b} spectral type versus T$_{\rm eff}$ relations, one expects a $\frac{\Delta {\rm T}_{\rm eff}}{\Delta {\rm M}_J}\sim310$\,K mag$^{-1}$ for T6 dwarfs, or a 0.13\,mag between DENIS0813 and SDSS1624+00 given the estimated difference of 40\,K. The consistency in absolute magnitude between these two objects suggests that DENIS0817 is unlikely to be an unresolved near-equal luminosity binary.

In retrospect, DENIS0817 was overlooked by previous searches in the DENIS and 2MASS surveys due to its relative proximity ($l = -14$, $b = 276$) to the galactic plane even though it is in a field sparse enough to allow for an efficient BD search without the crowding issues normally associated with moving object searches in the galactic plane.

This discovery serves as a reminder that, as suggested by recent LF estimates, the sample of known T dwarfs within 10 pc is largely incomplete. The CFBDS LF for isolated objects \citep{Reyle2009} indicates that there should be $6\pm1$ T0.5-T5.5 and $22_{-9}^{+13}$ T6 to T8 dwarfs within 10 pc of the Sun. This LF holds only for isolated and very-wide companions that can be identified with seeing limited observations (e.g. $\epsilon$ Indi Bab-like objects would be found but not Gl229b and SCR 1845-6357B). The number of known objects is about half of what is expected, suggesting that the census, even at moderate and high galactic latitudes, is still incomplete even for $J<16$ objects. Clearly, sensitivity alone cannot explain how objects such as DENIS0817 remained unidentified in catalogues for a decade. Near-infrared color-based searches alone are challenging for early-Ts as they have the same near-infrared colors as the largely more numerous M dwarfs. This problem is largely solved by including deep imaging in at least one filter blueward of $1\,\mu$m. Searches using non-simultaneous imaging are hampered by the number of false positives due to highly variable objects (flaring stars, novas, super-novas, AGB stars, etc) if they are bright when observing in near-infrared and fainter or absent when observing in the optical or far-red domain (e.g. 2MASS objects without visible counterpart). Multi-epoch far-red (Large Synoptic Survey Telescope,  Panoramic Survey Telescope \& Rapid Response System; LSST, Pan-STARRS), near-infrared (VISTA Variables in the Via Lactea; VVV) and mid-infrared (Wide-Field Infrared Survey Explorer; WISE) wide-field surveys are likely to be much more efficient in completing the local BD census \citep{Ivezic2008, Kaiser2002, Minniti2010, Mainzer2005}. 

The VVV is an ongoing ESO deep near-IR survey with VISTA at Paranal and will produce a multicolor $ZYJHK_{\rm s}$ atlas of the Milky Way bulge and inner disk. With its 5 passbands, high internal astrometric accuracy (15\,mas) and $\sim$$60$ epochs in $K_{\rm s}$ during the next 5 years, this survey will complete the census of the nearby L and T dwarfs at low galactic latitudes.

The proposed LSST will reach $z=23.3$ and $y=22.1$ ($5\sigma$) per visit, sufficient to detect objects similar to CFBDS0059 (T$_{\rm eff}\sim650$\,K, \citealt{Delorme2008}) out to $\sim$25\,pc. The expected accuracy for parallax measurements of faint objects over a 10-year baseline is on the order of a few mas and will be sufficient to determine the distance to all southern BDs within $25$\,pc to better than a few percents \citep{Ivezic2008}.

\begin{table}[!htbp]
\begin{center}
\caption{Properties of DENIS0817\label{tbl-1}}
\begin{tabular}{lc}
\hline
\multicolumn{2}{c}{DENIS}\\
Designation &DENIS J081730.0-615520\\
$J$&$13.371\pm0.09$\\
$K_s$&$13.207\pm0.20$\\
$J-K_s$&$0.164\pm0.22$\\
\multicolumn{2}{c}{2MASS}\\
Designation & 2MASS 08173001-6155158 \\
$J$    &$13.613\pm0.024$\\
$H$    & $13.526\pm0.031$\\
$K_s$  &$13.520\pm0.043$\\
$J-H$  &$0.087\pm0.039$\\
$H-K_s$&$0.006\pm0.053$\\
$J-K_s$& $0.093\pm0.049$\\
\multicolumn{2}{c}{Astrometry}\\
PM (RA) &$-0.336\pm0.054\arcsec$ yr$^{-1}$ \\
PM (DEC) &$ 1.095\pm0.041\arcsec$ yr$^{-1}$ \\
Parallax & $203\pm 13$ mas \\
distance & $4.9\pm0.3$ pc\\ 
M$_J$ (2MASS) & $15.15\pm0.14$\\
M$_J$ (MKO) $^a$&$14.85\pm0.14$\\
\multicolumn{2}{c}{Spectral indices$^b$}\\
CH$_4$-J & 0.306 (T6.68)\\
H$_2$O-H & 0.305 (T6.25)\\
CH$_4$-H & 0.321 (T6.36)\\
CH$_4$-K & 0.165 (T6.31)\\
K/J & 0.156          \\
K/H & 0.334          \\
KI[$1.243\mu$m]&$3.4\pm0.2$\,\ang\\
KI[$1.254\mu$m]&$7.1\pm0.3$\,\ang\\
Spectral type&T6\\

\hline\hline
\end{tabular}
\tablenotetext{a}{Determined from the M$_J$ (2MASS) using the \citet{Stephens2004} transformations. }
\tablenotetext{b}{Indices from \citet{Burgasser2006} }
\end{center}
\end{table}

\acknowledgments
EA thanks Lo\"ic Albert for thoughtful discussions about the DENIS versus 2MASS cross-match project. The authors thank Valentin D. Ivanov for his help in organizing the rapid spectroscopic follow-up at SOAR. SF acknowledges funding support from the ESO -- government of Chile Mixed Committee -- 2009, and also from GEMINI Conicyt grant No.32090014/2009. JB is supported by FONDECYT  No.1080086 and MIDEPLAN ICM Nucleus P07-021-F. This research has benefited from the M, L, and T dwarf compendium housed at \hbox{DwarfArchives.org} and maintained by Chris Gelino, Davy Kirkpatrick, and Adam Burgasser. This publication makes use of data products from the Two Micron All Sky Survey, which is a joint project of the University of Massachusetts and the Infrared Processing and Analysis Center/California Institute of Technology, funded by the National Aeronautics and Space Administration and the National Science Foundation.

\bibliographystyle{apj}


\end{document}